# Developing Bayesian Information Entropy-based Techniques for Spatially Explicit Model Assessment

Kostas Alexandridis and Bryan C. Pijanowski

*Abstract*— The aim of this paper is to explore and develop advanced spatial Bayesian assessment methods and techniques for land use modeling. The paper provides a comprehensive guide for assessing additional informational entropy value of model predictions at the spatially explicit domain of knowledge, and proposes a few alternative metrics and indicators for extracting higher-order information dynamics from simulation tournaments. A seven-county study area in South-Eastern Wisconsin (SEWI) has been used to simulate and assess the accuracy of historical land use changes (1963-1990) using artificial neural network simulations of the Land Transformation Model (LTM). The use of the analysis and the performance of the metrics helps: (a) understand and learn how well the model runs fits to different combinations of presence and absence of transitions in a landscape, not simply how well the model fits our given data; (b) derive (estimate) a theoretical accuracy that we would expect a model to assess under the presence of incomplete information and measurement; (c) understand the spatially explicit role and patterns of uncertainty in simulations and model estimations, by comparing results across simulation runs; (d) compare the significance or estimation contribution of transitional presence and absence (change versus no change) to model performance, and the contribution of the spatial drivers and variables to the explanatory value of our model; and (e) compare measurements of informational uncertainty at different scales of spatial resolution.

I. THE NEED FOR SPATIALLY COMPLEX STOCHASTIC MODELING ASSESSMENT

The aim of this approach is to explore and develop advanced spatial assessment methods and techniques for land use intelligent modeling. Traditional statistical accuracy assessment techniques, although essential for validating observed and historical land use changes, often fail to capture the stochastic character of the modeling dynamics. The research presented here provides a comprehensive guide for assessing additional informational entropy value of model predictions at the spatially explicit domain of knowledge. It proposes a few alternative metrics and indicators that encapsulate the ability of the modeler to extract higher-order information dynamics from simulation experiments. The term *information entropy*, originates from the information-theoretic concept of entropy, conceived by Claude Shannon on his famous two articles of 1948 in Bell System Technical Journal [1], and expanded later in his book "Mathematical Theory of communication" [2]. Since the mid-20th century, the field of information theory has experienced an unprecedented development, especially following the expansion of computer science in almost every scientific field and discipline. The concept of entropy in information systems theory allow us to allocate quantitative measurements of uncertainty contained within a random event (or a variable describing it) or a signal representing a process [3, 4].

The literature on assessing spatially explicit models of land use change has made substantial steps during the last few years. Many of the metrics and assessment techniques in the past have been treating land use predictions as complex signals, and models themselves often are treated as measurement instruments, not different from signal-measurement devise assessment in physical experiments [5]. Spatially explicit methods and assessment techniques are used in many remote sensing applications [6]; wildlife habitat models [7]; predicting presence, abundance and spatial distribution of populations in nature [8]; analyzing the availability and management of natural resources [9, 10]. In a more theoretical level of analysis, spatially explicit methods of model assessment have been used for testing hypotheses in landscape ecological models [11, 12]; address statistical issues of uncertainty in modeling [10]; or analyze landscape-specific characteristics and spatial distributions [13].

Methodologies and techniques as the ones referenced above often maintain and preserve traditional statistical approaches to modeling assessment. Most likely, they test the limitations and assumptions of statistical techniques originally designed for analyzing data and variables that do not exhibit spatially explicit variation. The majority of studies where

Manuscript received June 2008. This work was supported in part by a Purdue Research Foundation (PRF) Fellowship, the US National Science Foundation (Grant WCR:0233648), the EPA STAR Biological Classification Program, the Great Lakes Fisheries Trust, and the Purdue University Department of Forestry and Natural Resources.

K. *Alexandridis* is a Research Scientist (Regional Futures Analyst) with the Commonwealth Scientific and Industrial Research Organization (CSIRO), Division of Sustainable Ecosystems, CSIRO Davies Laboratory, University Drive, Douglas, QLD 4814, Australia (phone: +61 7 4753 8630; fax: +61 7 4753 8650; e-mail: Kostas.Alexandridis@csiro.au).

B. C. *Pijanowski*, is an Associate Professor with Purdue University, Department of Forestry and Natural Resources, West Lafayette, IN, USA (e-mail: bpijanow@purdue.edu).

spatially explicit methodologies are used tend to involve relatively simple or linear statistical analyses [14, 15]. While in the recent years assessment of modeling complexity has been an issue of analysis [16, 17], has yet to include spatial complexity and assessment of stochasticity as essential elements of evaluation and analysis. Spatial complexity by itself is not often enough to fully describe and represent the complex system dynamics of coupled human and natural systems. The introduction of spatial complexity in advanced dynamic modeling environments requires the involvement of stochasticity as an essential element of the modeling approach. It rests between traditional spatial assessment and game-theoretic approaches to modeling. The level of uncertainty and incomplete information embedded on the components of a coupled human-biophysical system often necessitates the introduction of stochasticity as a measurable dimension of complexity [18, 19]. Stochastic modeling is widely introduced in modeling complex natural and ecological phenomena [20], population dynamics [21], spatial landscape dynamics [22], intelligence learning and knowledge-based systems [23], economic and utility modeling [24, 25], decision-making, Bayesian and Markov modeling [26, 27] and many other associated fields in science and engineering applications.

A natural extension of the related techniques and methodologies is the development and introduction of spatially explicit, stochastic methods of accuracy assessment for intelligent modeling. In recent years, methods, techniques, and measures of informational entropy exceeded the single dimensionality of traditional statistical techniques (i.e., measuring uncertainty on single random events or variables) and begun analyzing multi-dimensional signals. The concept of spatial entropy [28, 29] presents analysis of informational entropy patterns in two-dimensional spatial systems. Within these lines, the remaining of the paper introduces some alternative metrics that aim to assist and enhance the power of our inferential mechanisms in modeling such systems.

## II. A CASE-STUDY: ANN SIMULATIONS IN SOUTH-EASTERN WISCONSIN REGION

The study is based on modeling historical urban spatial dynamics using Artificial Neural Network (ANN) simulations for a large spatial region of South-Eastern Wisconsin (SEWI) in the Midwestern region of U.S. The details of the simulation can be found in a recent paper by Pijanowski et al. [30], where the modeling dynamics and a comprehensive description of the LTM modeling mechanism and experimental design is deployed. Description of the LTM model is also provided in Pijanowski et al. [31, 32].

### A. Sampling Methodology

The project area involves a seven-county region in the South-Eastern Wisconsin (SEWI) region, and includes the city of Milwaukee and its wider suburban area [33]. The land use changes that occurred in the SEWI region during the period 1963-1990 is considerable. Most of the urban growth has taken place in the suburban metropolitan Milwaukee region, and the areas around medium and large cities in the region (Fig. 1). The county of Waukesha, in the west side of the city of Milwaukee has absorbed the majority of suburban changes, but important urban and suburban changes have occurred in the remaining counties both at the North (Washington and Ozaukee counties) and South (Walworth, Racine and Kenosha counties) of the city of Milwaukee.

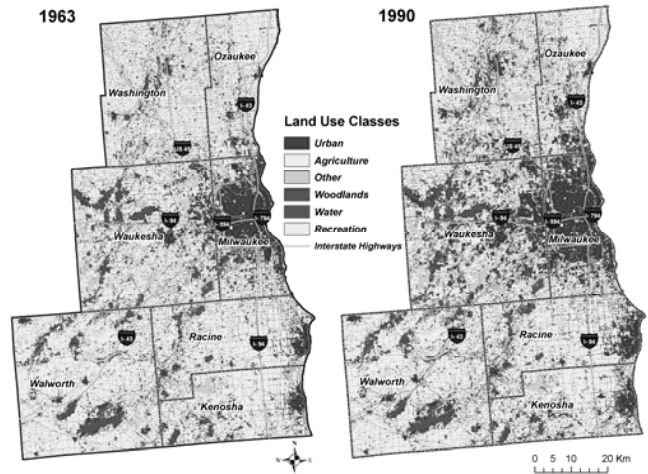

Fig. 1: Land use changes in the SEWI region, 1963-1990.

The large size of the area under study, and the ability to perform extensive training and learning simulations using the LTM model, makes computationally impossible to simulate the entire region as a whole. Instead, a comprehensive sampling methodology has been implemented. The regional extent of the SEWI area has been divided into equal-area square boxes of 2.5 square kilometers (or 6889 cells of 30 m$^2$ resolution). The square sampling boxes vary on both number of cells that experienced urban change during the 1963-1990 period, and the amount of exclusionary land zones (urban zones in 1963, paved roads, water bodies, protected areas, etc.). Both parameters affect the modeling performance and the ability to assess comparatively the accuracy of the modeling predictions. Thus, random sampling scheme has been implemented for this modeling exercise ensuring comparative assessment of the quantities and spatial patterns of land use change in the region. First, the regional sampling boxes has been ranked and classified using a combined index of both proportion of urban change and proportion of exclusionary zone within the sampling box. The yielded combined ranking index takes account of both changes within the sampled boxes and represents the ratio between the percentage of urban change and the percentage of variation in exclusionary zone areas across the sampled boxes[1]:

$$I_s = \frac{\%\Delta(urban)}{\%\Delta(exclusionary)} \quad (1)$$

where $s$ = number of area sampling boxes in the landscape.

---

[1] In LTM model, an exclusionary zone is defined as the map area where model pattern training and simulation are not implemented, i.e., areas with no suitability for transitional change. Examples of these zones include map areas

From the continuous sampling index values derived from the previous step, two threshold values of the sampling index have been used to define three classification index regions for random sampling. The sampling boxes have been assigned into three sequential classification pool groups (group A, B and C), according to the following rules (thresholds):

$$\text{Sampling Pool Group}: \begin{cases} A & if \quad I_s \geq 0 \\ B & if \quad I_s \geq 1/2 \\ C & if \quad I_s \geq 1 \end{cases} \quad (2)$$

The sampling pool classification in equation (2) follows a nested hierarchical scheme, that is, the prospective sampling pool of each consequent group is contained in the previous one (i.e., sampling pool for group C is fully contained in group B's sampling pool, and sampling pool for group B is fully contained within group A's sampling pool). Such classification scheme allow the testing of the effects of increased exclusionary zone area to the model performance in the simulations. Sampling pool group A contains all boxes in the sampling region. Sampling pool group B contains only sampling boxes that have no more than double the percentage of exclusionary area than the percentage of urban change area. Finally, sampling pool group C contains only the sampling boxes that have no more than equal or more percentage of urban change area than exclusionary zone area within them. The members (sampling boxes) of each sampling pool group (A, B, and C) have been ranked and assigned to 30th quintiles according to their ascending proportion of urban change within the sample box. From each 30-tile, one sampling box has been randomly selected using a random number generator algorithm. The seed of the random number generator has been renewed before each sampling operation. The final outcome of the random sampling procedure, was three sampling groups (varying on the ascending ratio of urban to exclusionary zone area), containing thirty 2.5 square kilometer sampling boxes each (varying on the percent of urban change).

The sampled boxes for area groups A, B and C are shown in the following Fig. 2.

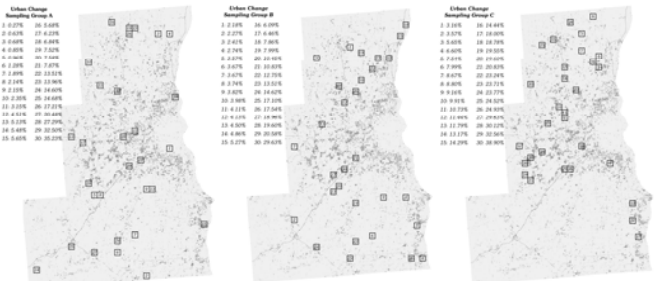

Fig. 2: SEWI random sampling groups A, B and C.

### B. Simulation Modeling Parameterization

The LTM model requires three levels of parameterization: (a) the simulation drivers of land use change; (b) the training and testing neural network pattern creation; and, (c) the network simulation parameter definition. Details on the theoretical neural network simulation parameterization of the LTM model are reported in the literature by Pijanowski et al. [31, 32]. Explicit description of the modeling enterprise in the SEWI region are also reported in the Pijanowski et al. [30] paper. In short, eight simulation distance drivers of land use change has been used to parameterize the LTM simulations (urban land in 1963, historical urban centers in 1900, rivers, lakes and water bodies, highways, state and local roads, and lake Michigan). The simulation model uses every other cell (50% of the cells) as neural network training pattern, and the entire region for network model testing. Finally, the network is trained for 500,000 training cycles, by resetting and iterating the network node's weight configuration every 100 training cycles, and outputting the network node structure and the mean square error of the network convergence every 100 cycles in a file. Thus for each of the 90 sampled boxes, the simulation output a total of 5,000 network files and MSE values for a grand total of 450,000 simulation result files. For simplicity of presentation, in each of the sampled boxes, forty-four of these network output files are selected for visualization of the results. Due to the nature of the neural network learning dynamics, learning patterns follow a negative exponential increase through training iterations. Thus, a negative exponential visualization scale has been chosen to visualize the results (more frequent samples in lower network training cycles, less frequent samples in higher training cycles).

The simulation results for urban change predictions in 1990 are assessed against historical land use changes in 1990 from existing data provided by the Southeastern Wisconsin Regional Planning Commission [33].

### III. METHODOLOGY AND RESULTS OF THE SIMULATION ACCURACY ASSESSMENT

The paper by Pijanowski et al. [30], reports three relative conventional statistical metrics for the quantitative accuracy assessment of the model performance. Namely, the *percent correct* metric (PCM), the *Kappa* metric (K), and the *area under the receiver operator characteristic curve* (AROC) metric. The PCM metric is a simple proportional measure of comparison, while the K and AROC metrics take into account the confusion matrix and the omission and commission errors of the simulation. In addition to these three conventional metrics[2], two more alternative metrics are presented here. Namely, the *Bayesian predictive value of a positive and negative classification* (PPV and NPV) metrics, and the *Bayesian conversion factor* ($C_b$) metric. These alternative metrics measure a stochastic level of information entropy in the simulated land use change system. They represent

---

that already undertook transitions before the simulations' initialization; road and transportation system extends; preserved natural areas, etc.

[2] The discussion involving these three metrics is largely omitted from this paper. The reader can consult the relative literature, and the Pijanowski et al.

different aspects or dimensions of the predictive value of information that is embedded in the simulation model results, and thus, can enhance our understanding of both simulation dynamics, and the dynamics of the land use change system.

*A. Basic Definitions*

The notion of spatial accuracy assessment utilizes three major assumptions. The first assumption has to do with the underlying process in hand. In any given landscape, two theoretical observers (e.g., a simulation model and an observed historical map, or a simulation model and another simulation model, or an observed historical map and an alternative historical map) are assumed to observe properties of the same underlying process (the "real" land use change). The second assumption has to do with the observers themselves. They are assumed to face a theoretical level of uncertainty (regardless the degree of, small or large). A simulation model is facing uncertainty on its predicted landscape as a part of the problem formulation (and thus a trivial assumption), but observed historical landscapes are also subjects to an implicit degree of uncertainty (i.e., measurement errors, remote sensing classification errors, etc.). These degrees of uncertainty are not necessarily equal between the two observers. The third assumption involves the assessment process itself. It assumes that the two observers acquire their observations (classification) independently from each other. In other words, the historically observed land use map and the simulation results are independent (or, the modeling predictions are not a function of the real change in the maps). The independence assumption is easily to assume in the case of assessing a simulated and an observed landscape, but it becomes nontrivial when non-parametric analysis is used to compare two modeled landscapes, in cases where the same model with different configuration is used.

A parametric approximation of spatial accuracy assessment is based on the notion of a confusion matrix [34, 35] shown in Table 1. For binary land use changes (i.e., presence-absence of transition), the confusion matrix is a 2×2 square matrix with exhaustive, and mutually exclusive elements.

Table 1: Theoretical confusion matrix for binary spatial accuracy assessment. Simulated versus historical land use change (TN: true negative, TP: true positive, FN: false negative, FP: false positive, SN: simulated negative, SP: simulated positive, RN: real negative, RP: real positive, GT: grand total).

|  | Observed 0 | Observed 1 |  |
|---|---|---|---|
| Simulated 0 | TN | FN | SN |
| Simulated 1 | FP | TP | SP |
|  | RN | RP | **GT** |

(2007) paper for more information. Only their value and performance for the simulation experiments in the SEWI region will be report.

A nonparametric approximation of spatial accuracy assessment employs the use of the confusion matrix in somewhat more complex forms. It assesses the sensitivity coefficient as the observed fraction of agreement between the two assessed landscapes, or, in other words, the probability of correctly predicting a transition when this transition actually occurred in the observed historical data. Symbolically (S=simulated, R=real),

$$Sensitivity = \frac{TP}{TP+FN} = p(S=1|R=1) \quad (3)$$

Similarly, the specificity coefficient in equation (4) represents the observed fraction of agreement between two assessed maps, or, in other words, the probability of correctly predicting an absence of transition, when this transition is actually absent from the historically observed data. Symbolically,

$$1-Specificity = \frac{TN}{TN+FP} = p(S=0|R=0) \quad (4)$$

A theoretical *perfect* agreement between the two observers would require that,

$$p(S=1|R=1) = p(S=0|R=0)$$
$$or, \; Sensitivity = 1-Specificity \quad (5)$$

The degree of deviation from the rule as defined in equation (5), represents the degree of deviation from a perfect agreement between the two classifications, or the degree of disagreement between a modeled (simulated) and an observed (historical) landscape transition. The binary character of the classification schemes requires the two transition classifications to be exhaustive and mutually exclusive. The theory of statistical probabilities suggests that a random (fully uncertain) classification between the probabilities denoted by sensitivity and specificity coefficients would be:

$$Sensitivity = Specificity = \frac{1}{2} \quad (6)$$

In other words, for each classification threshold (e.g., amount of urban change) in our assessment, a given cell has an equal (prior) chance (50%) to undergo a land use change transition, not unlike the tossing of a coin.

*B. Bayesian Predictive Value of Positive and Negative Classification metric (PPV / NPV)*

*1) Diagnostic Odds Ratio (DOR)*

From the definitions of sensitivity and specificity in the previous session, we can compute the *likelihood ratio* metric [36, 37]. In a binary (Boolean) classification scheme, there are two forms of likelihood ratios: the *likelihood ratio of a positive classification* (LR+), and the *likelihood ratio of a positive classification* (LR-). The likelihood ratios are connected with the levels of sensitivity and specificity directly [38]:

$$LR+ = \frac{sensitivity}{1-specificity} \quad and \quad LR- = \frac{1-sensitivity}{specificity} \quad (7)$$

The likelihood ratios obtained for a binary classification can be used to compute the value of an index for diagnostic inference, namely, the *diagnostic odds ratio* (DOR) index. The DOR represents simply the ratio of the positive to the negative likelihoods:

$$DOR = \frac{LR+}{LR-} = \frac{sensitivity \cdot specificity}{(1-specificity) \cdot (1-sensitivity)} \quad (8)$$

The DOR can be interpreted as an unrestricted measure of the classification accuracy [38], but suffers from serious limitations, since both LR+ and LR- are sensitive to the threshold value (cut-off point) of the classification [39]. Thus, DOR can be used as a measure of the classification accuracy in cases where, (a) the threshold value of the binary classification is somewhat balanced (around 0.5), or; (b) when comparing classification schemes that have the same threshold value (e.g., in the case of simulation runs that are unbalanced but face similar threshold values). In the case of the SEWI region simulation runs, the DOR can be used to compare classification performance across training cycles (same areas, and same classification thresholds), but not across area groups or different simulation boxes. The results shown in Fig. 3a, signify the importance of pattern learning (training) process of improving the classification accuracy in the SEWI region experimental simulations.

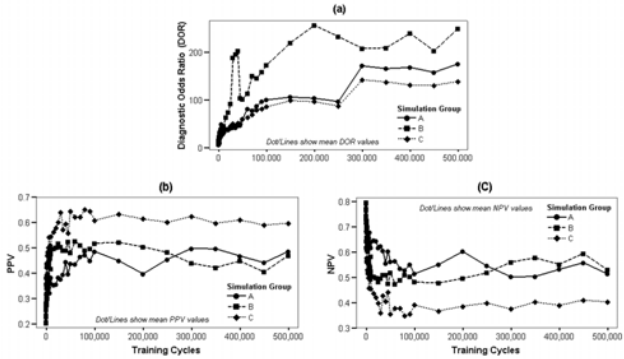

Fig. 3: (a) Diagnostic Odds Ratio (DOR) index across LTM training cycles in the SEWI region; (b) Bayes probability of change given a positive classification (PPV); (c) Bayes probability of change given a negative classification (NPV) metric results by simulation group in the SEWI region.

*2) Bayesian Predictive Values*

In place of the simple and practically limited DOR index to assess the robust spatial model accuracy, a Bayesian framework of assessment can be used. It uses the likelihood ratios (LR+ and LR-), to estimate a posterior probability classification based on the information embedded in the dataset. Strictly speaking, the model accuracy obtained by the confusion matrix (and consequently the sensitivity and specificity values), represents a *prior* probabilistic assessment of the model's accuracy. This assessment is subject to the threshold value of the classification scheme. Obtaining a classification scheme that is robust enough to allow us to estimate model accuracy for a range of thresholds, requires the computation of the conditional estimates [40]. This represents a *posterior* probabilistic assessment of the model's accuracy, and can be achieved using Bayes' Theorem. Computing the posterior Bayes probabilities for a positive and negative classification can be achieved using a general equation form:

$$PPV = p(x_+ \mid c) = \frac{p(x_+) \cdot p(c)}{p(x_+) \cdot p(c) + p(x_-) \cdot p(1-c)} \quad (9)$$

and,

$$NPV = p(x_- \mid 1-c) = \frac{p(x_-) \cdot p(1-c)}{p(x_-) \cdot p(1-c) + p(x_+) \cdot p(c)} \quad (10)$$

where,
PPV: the Bayes predictive value of a positive classification metric; NPV: the Bayes predictive value of a negative classification metric; $x_+, x_-$: the positive and negative values of the classification, and,; $c$: the *prevalence* threshold for which a value is positive if it is larger or equal from (computed using a ML nonparametric estimation).

The PPV and NPV values can be computed from the *sensitivity* and *specificity* values (and thus from the confusion matrix) as follows:

$$PPV = \frac{sens \cdot prev}{sens \cdot prev + (1-sens) \cdot (1-prev)} \quad (11)$$

and,

$$NPV = \frac{(1-spec) \cdot (1-prev)}{(1-spec) \cdot (1-prev) + sens \cdot prev} \quad (12)$$

The results for the PPV and NPV metrics obtained for the SEWI region and the three simulation area groups are shown in the following Fig. 3b,c. Simulation area group C has consistently the higher PPV and the lowest NPV throughout the training exercise, a fact that signifies a higher model performance level than the ones achieved by simulation area groups A and B.

Measuring and treating PPV and NPV as separate metrics of model performance is a rather trivial operation, and it is not a very useful or informational tool in assessing spatial model accuracy. However, by combining the PPV and NPV metrics into a single graph, we can illustrate the dominance relationships and dynamics over an expected prevalence threshold value (i.e., *prevalence*=0.5, denoting an uninformative prior for the Bayesian classification). Fig. 4 shows the dominance relationships between PPV and NPV for increasing LTM training cycles. In simulation area group A, the model accuracy is based mainly on the dominant negative classification (although this dominance fades over the training process). The accuracy in simulation area group B is based on an unstable equilibrium between positive and negative classification (especially between 20,000 and 250,000 training cycles), although the overall accuracy is still supported by a dominant negative classification scheme. The model accuracy in simulation area C depends on a more desired classification scheme, since after the first 10,000 cycles model accuracy depends consistently on a positive classification.

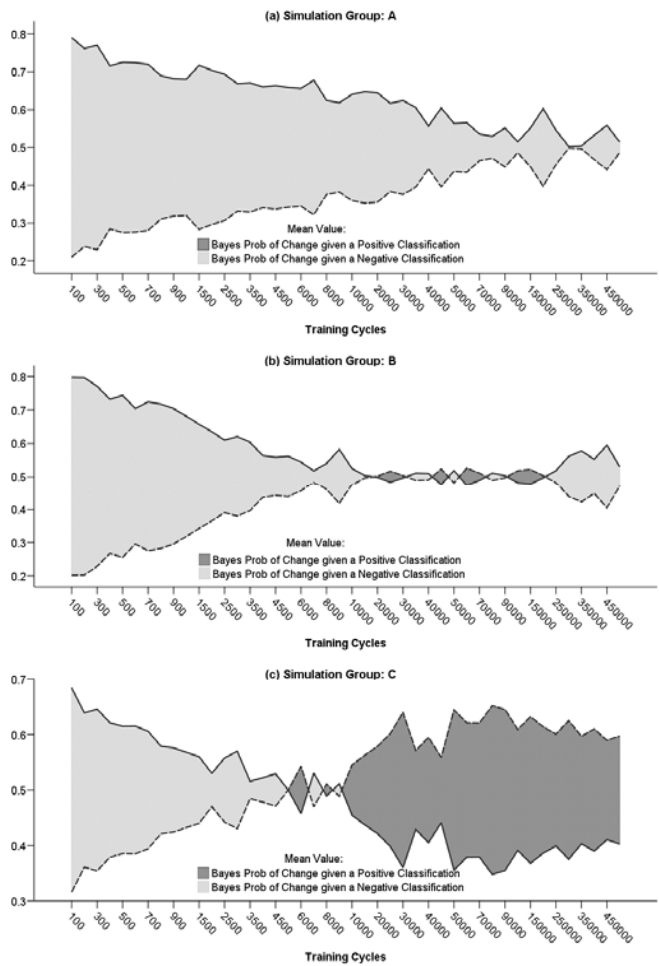

Fig. 4: Dominance relations between Bayes PPV and NPV metrics: (a) area group A; (b) area group B; (c) area group C.

The analysis of the latter results is based on an expected, i.e., balanced prevalence threshold. In reality, the relationship between Bayesian predictive values and prevalence is non-linear and it is defined by the posterior Bayesian estimator properties, namely the posterior density estimation [36, 41].

To understand the role of the posterior Bayesian estimation, a theoretical problem formulation is provided in Fig. 5. Part (a) of the figure provides a hypothetical prior density estimation of a binary classification scheme across a continuous range of classification thresholds (prevalence). For a given transitional change (e.g., presence of land use change), the prevalence threshold ranges from zero (purely negative) to one (purely positive). The left density curve represents the absence of a transition (negative classification), while the right density curve represents the presence of a transition (positive classification). As explained in the first section, when we lack any additional information about the classification threshold, the best uncertain choice (maximum entropy classification), is to assume an equal probability between the two classes (present, absent). In most of the cases involving spatial accuracy assessment, an uncertain prior is the best choice. Unlike the ROC curve method, where accuracy is assessed using a nonparametric estimation (without the use of a distribution function), the Bayesian estimation is based in a parametric assessment of the classification accuracy (or, at least a semiparametric assessment). In such an uncertain classification, we can vary only the spread of the distribution (i.e., the width of the density distribution) for each of the classes, but not the location of the threshold. As a consequence, the amount and proportions of the false negative (FN) and false positive (FP) allocations are affected only by the difference on the mean value of each of the transitions to the threshold. The more this difference is positive, the more likely it is for the transition to be present, while the more the difference is negative, the more likely it is for the transition to be absent.

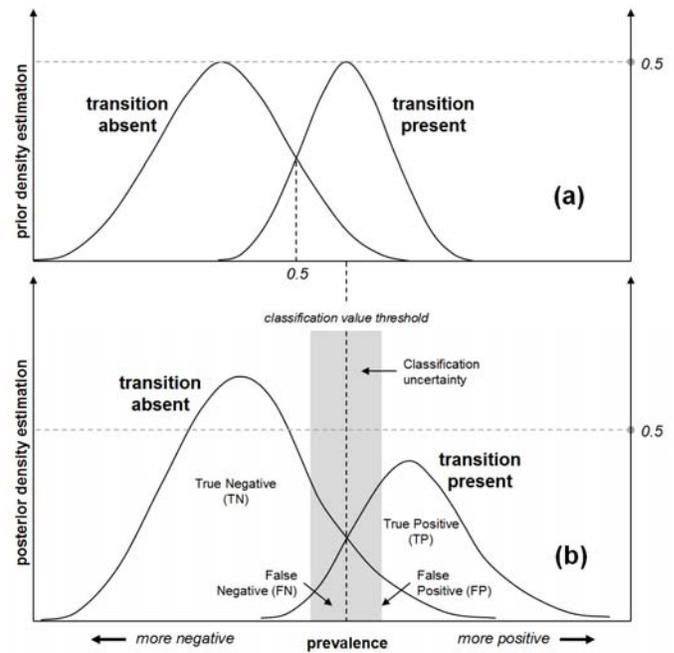

Fig. 5: Properties of the Bayesian estimation in binary classification scheme: (a) prior density estimation; (b) posterior density estimation.

Bayesian estimation allows us to estimate the probability densities of the classifications by adjusting the "true" height and "true" width of the density distributions. In Fig. 5b, the changes in the density distributions for the threshold classes shifts the threshold prevalence value disproportional to the size and spread of each of the distributions. The posterior Bayesian density estimates allow us to evaluate the mean and variance of a new, "informative" prevalence threshold (shown with dotted line and shaded areas in Fig. 5b).

In the SEWI region, the relationship between prevalence and the level of the PPV/NPV is shown in Fig. 6. The y-axis of the graph represents the prevalence level (classification threshold), while the x-axis represents the level of the predictive value (PPV or NPV). The points that belong to the PPV and NPV are color-coded. The data points correspond to all sampled simulation runs (44 sampled training cycles for each of the 90 boxes in groups A, B and C, a total of 3,960 simulation run results).

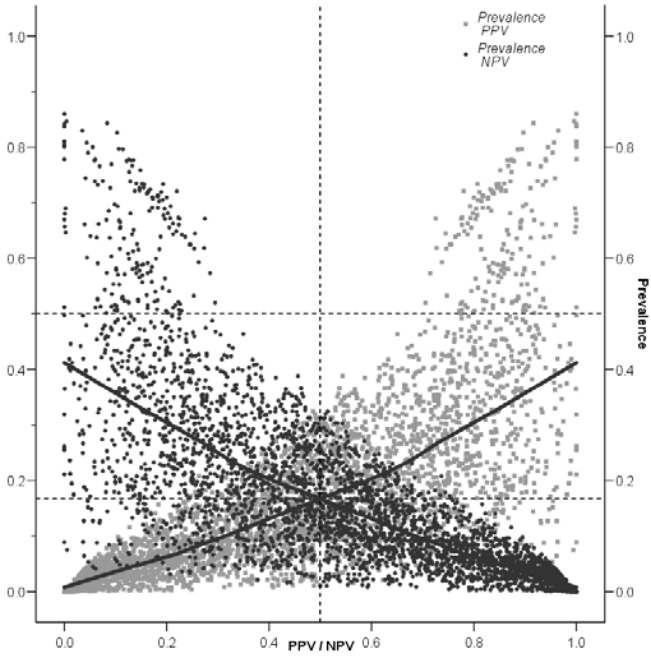

Fig. 6: Relationship between prevalence (y-axis) and the Bayes predictive values, PPV and NPV (x-axis). The solid lines represent the Epanechnikov kernel density estimation for PPV and NPV. The dotted reference lines identify the difference between expected and predicted prevalence thresholds.

We can perform a nonparametric estimation of the probability density function in the data, by using a kernel density estimator. The solid lines in Fig. 6 represent the results of the *Epanechnikov* stochastic kernel estimation [42]. The general equation of the kernel density function is [43, 44]:

$$\hat{f}_K(x) = \frac{1}{Nh}\sum_{i=1}^{N} K\left(\frac{x-X_i}{h}\right) \quad (13)$$

where,

$\hat{f}_K$: an unknown continuous probability density function; $h$: a smoothing parameter; $K(z)$: a symmetric kernel function, and; $N$: the total number of independent observations of a random sample $X_N$.

The equation for the Epanechnikov kernel density function is [43]:

$$K(z) = \begin{cases} \frac{3}{4\sqrt{5}}(1-\frac{1}{5}z^2) & if \quad -\sqrt{5} \leq z \leq \sqrt{5} \\ 0 & otherwise \end{cases} \quad (14)$$

The choice of the Epanechnikov kernel density estimator is based on the high efficiency on minimizing the *asymptotic mean integrated square errors*, AMISE [45, 46], and it is often used in Neural Network computational learning [47].

In the SEWI region data, the underlying question that the analysis attempts to address is for which prevalence threshold value the "true" predictive value (and accuracy) of the modeled transitional classification becomes equal to the "true" absence of such transaction? Graphically, the solution can be found by varying the height of the y-axis reference line (horizontal dotted lines in Fig. 6) over a fixed level of predictive value, where $PPV = NPV = 0.5$ (vertical dotted line). The y-axis coordinate for which the two kernel density estimated lines meet represents the prevalence threshold that maximizes the posterior probability of our model accuracy predictions.

Mathematically, the optimal prevalence threshold of the posterior probability distribution exists where:

$$\hat{f}_K(x_+) = \hat{f}_K(x_-) \quad (15)$$

The difference between the prior and posterior estimation is shown in the vertical distance between the y-axis reference lines at the 0.5 prevalence threshold and the one at the meeting point of the two kernel density functions (~0.172 in the entire SEWI regions' simulation data). The posterior estimation allows us to threshold at a lower classification level, and thus enhancing the accuracy of our predictions.

*3) Bayesian Convergence Factor metric ($C_b$)*

It is possible to derive an alternative accuracy metric that combines the two Bayesian predictive values, PPV and NPV in a single, unified coefficient. The use of such a coefficient to measure classification and model accuracy is that allows us to estimate not only a unique prevalence threshold, but also an optimal prevalence region for which our estimated accuracy is high for both positive and negative classifications. The analysis provided in the previous paragraph in the case of PPV and NPV metrics depends mainly on the choice of the kernel density estimation function and the continuous interval *bandwidth* used [43, 48], or any other probability density function used for estimation. A unified Bayesian coefficient that measures the level of convergence between positive and negative predictive values permits us to derive a more robust prevalence region that tends to smooth the effect of density estimation selection. In other words, it provides us with a more global measure of model and classification assessment.

We can call this coefficient *Bayes convergence factor*, $C_b$. A simple form of the factor can be defined as:

$$C_b = \begin{cases} 1-(PPV-NPV) & if \quad PPV \geq NPV \\ 1-(NPV-PPV) & if \quad PPV < NPV \end{cases} \quad (16)$$

A higher level of the Bayes convergence factor thus denotes higher probability of convergence between a positive and a negative predictive value or probabilities of change. Because of the probability properties of such a coefficient, and the fact that always $PPV + NPV \leq 1$ (the probability of change cannot exceed 1.0), the range of the $C_b$ coefficient will be: $0 \leq C_b \leq 1.0$. This simple form of the Bayes convergence factor is shown in the theoretical curve $C_b(A)$ of Fig. 7a. We can see that the allocation of the positive and negative classification probabilities in the $C_b$ function represents a form of a *triangular* density function with minimum value of zero, maximum value of 1.0, and mean value of 0.5. A triangular density function provides a minimal amount of information about the relationship, configuration and pattern between the positive and negative predictive values in a model. As shown in Fig. 7a, these predictive values by themselves may be better represented by non-linear relationships (e.g., kernel density estimators). Thus, a better convergence factor can be found that reflects a degree of nonlinearity in the modeling classification assessment.

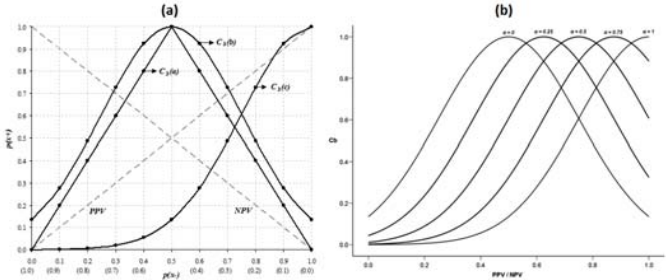

Fig. 7: (a) Theoretical distribution density functions for the Bayes convergence metric (Cb): (i) triangular; (ii) adjusted normal; (iii) asymmetric normal. (b) Variations of the asymmetry parameter, α, in the expected normal form of the Bayes convergence factor.

An alternative form of the Bayes convergence factor can be symbolically calculated using a *Normal density distribution function*, adjusted to a continuous scale between 0 and 1.0. The equation of the Normal density function is,

$$\hat{f}_N(x,\mu,\sigma) = \frac{1}{\sqrt{2\pi}\sigma} \cdot e^{-\left(\frac{(x-\mu)^2}{2\sigma^2}\right)} \tag{17}$$

For a Normal distribution with $x = 0$ and $\sigma = 0.5$, we can model the behavior of the mean value, by setting,

$$\mu = PPV - NPV \tag{18}$$

and thus,

$$\hat{f}_N(0, PPV-NPV, 0.5) = \frac{1}{\sqrt{2\pi}(0.5)} \cdot e^{-\left(\frac{(0-(PPV-NPV))^2}{2(0.5)^2}\right)} \tag{19}$$

$$= 0.797885 \cdot e^{-2(PPV-NPV)^2}$$

We can adjust for the coefficient scale (0 to 1.0), by multiplying the previous equation by a normalization factor,

$$\frac{1}{\sqrt{2\pi}\sigma} = 0.797885 \tag{20}$$

The *adjusted Normal* form of the Bayes convergence factor, can expressed as:

$$C_b = e^{-2(PPV-NPV)^2} \tag{21}$$

The adjusted normal density distribution function of the $C_b$ coefficient can be seen in the curve $C_b$ (B) of Fig. 7a, and in our data can be estimated by a Normal or Epanechnikov kernel density function.

The previous two forms of the $C_b$ metric assume implicitly that the combined effect of the positive and negative classification process in our model is symmetric toward achieving a better model (and classification) accuracy. It is appropriate for modeling changes where the presence of a transition implies the absence of a negative transition. In many spatial modeling processes simulating binary change that implicit assumption cannot be made easily. For example, a model (such as LTM) that simulates land use change is parameterized and learns to recognize patterns on drivers of change related to a positive land use transition effect only. Model training and testing based on drivers of transitional presence, do not necessarily convey information on the probability of absence of such a transition, as it is likely that other or additional drivers of the absence of the transition may be in effect over an ensemble of landscapes. Consequently, we can derive a better form of the Bayes conversion function by assuming a biased or asymmetric join distribution among the predictive value of positive and negative classification. Such an asymmetry would favor more positive than negative classifications, assuming that the model learns more about the transitional patterns from a combination of a high positive and low negative predictive value, rather than from a high negative and low positive predictive value (since the sum of the predictive values equals 1). The later is especially important in estimating empirical distributions derived from unbiased real-world data, such as in the SEWI case study. The amount of area that undertakes urban land use transition in the data is considerably less than the amount of area that observes an absence of such transition, and implementing an asymmetric Bayesian prior distribution would assign more weight in the positive (presence of transition) than in the negative (absence of transition) land areas.

We can formulate such a conversion function from modifying the mean central tendency of the previous form, $C_b$ (B). In other words, by simulating a different mean for the adjusted Normal distribution function. We can call this form, *adjusted asymmetric Normal density distribution*, and for the same numerical parameters, $x = 0$, and $\sigma = 0.5$, we can simulate the behavior of the mean value,

$$\mu' = \alpha - (PPV - NPV) \tag{22}$$

where, $\alpha$ is the degree of asymmetry of our distribution ($0 \le \alpha \le 1.0$). In other words, the parameter $\alpha$ denotes the degree of bias in terms of a theoretical *least-cost function*, or the relative informational balance in our model from a positive to negative predictive value.

The new asymmetric normal distribution will be,

$$\hat{f}_N(0, \alpha-(PPV-NPV), 0.5) = \frac{1}{\sqrt{2\pi}(0.5)} \cdot e^{-\left(\frac{(0-(\alpha-(PPV-NPV)))^2}{2(0.5)^2}\right)} \tag{23}$$

$$= 0.797885 \cdot e^{-2(PPV-NPV-\alpha)^2}$$

and, after adjusting for scale normalization, the final Bayes convergence factor, will be,

$$C_b = e^{-2(PPV-NPV-\alpha)^2} \tag{24}$$

For varying levels of the parameter $\alpha$, the shape of the latter convergence factor is shown in Fig. 7b. For $a = 0$, the equation yields the *symmetric normal* form of the convergence factor (i.e., shape $C_b$ (B) in Fig. 7a), while, for $a = 1.0$, the equation yields a *full asymmetric normal* form of the convergence factor (i.e., shape $C_b$ (C) in Fig. 7a). In an experimental dataset, any form of asymmetric normal distribution form of $C_b$ (i.e., for any parameter $\alpha$) can be estimated by a Normal of Epanechnikov kernel distribution function.

The results of the empirical data obtained for the SEWI region simulation runs for the varying degree of asymmetry in estimating the Bayes Convergence Factor, $C_b$, are shown in Fig. 8a. We can see that a somewhat moderate level of asymmetry ($\alpha = 0.25$) performs consistently better throughout the entire model learning process (training cycles), despite the fact that at the 500,000 cycles training cycle level, the Bayes Convergence Factor with $\alpha = 0.5$ performs slightly better. Thus, there is evidence in the SEWI simulation runs that a level of asymmetry in the composition of positive and negative predictive value of our model exists, and thus should be incorporated into our spatial accuracy assessment.

Table 2: Estimated values for the location (μ) and scale (σ) parameters of the empirical asymmetric Bayes Convergence Factor in the SEWI area.

| Selection Level | Bayes Conversion Factor Assymetry | Predicted Values for Bayes Convergence Factor | |
|---|---|---|---|
| | | Estimated location of Normal Distribution (Mean) | Estimated scale of Normal Distribution (SD) |
| across all training cycles | a=0 | .5985 | .3027 |
| | a=0.25 | .5571 | .3411 |
| | a=0.5 | .4764 | .3740 |
| | a=0.75 | .3651 | .3685 |
| | a=1 | .2443 | .3092 |
| after 500,000 training cycles | a=0 | .5067 | .3114 |
| | a=0.25 | .5537 | .3416 |
| | a=0.5 | .5608 | .3861 |
| | a=0.75 | .5019 | .4128 |
| | a=1 | .3834 | .3772 |

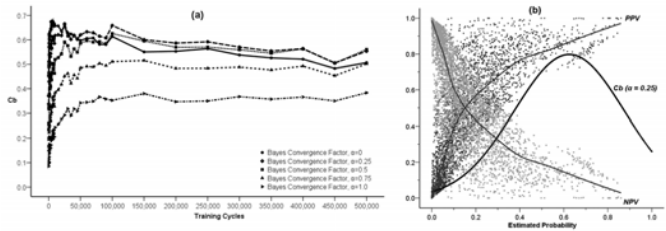

Fig. 8: (a) Estimated mean Bayes convergence factor for varying levels of the asymmetry coefficient in the SEWI region; (b) Estimated probabilities of transition from the empirical values of the Bayes convergence factor (α=0.25). Data points represent the estimated PPV and NPV values in the SEWI region (for all simulation boxes' sampled training cycles).

Beyond any visual inspection and inference of our results, it is possible to derive quantitative estimates of the dominance of a level of asymmetry present in our simulation runs. As can be seen in Fig. 8b we can estimate the expected probabilities of transitions, subject to the observed empirical values of transitions present in our simulation data. When all the simulation runs results for the entire SEWI region are examined with respect to their respective observed predictive values, we can estimate such an empirical probability distribution, as a function of an estimated "true" mean (location parameter) and standard deviation (scale parameter) of each of the forms of Bayes Convergence Factor, $f_N(\hat{\mu}_{C_b}, \hat{\sigma}_{C_b})$, using a maximum likelihood estimation (*ML*) method. The results of such estimation for the varying degree of asymmetry in the Bayes convergence factor in the SEWI data are shown in Table 2. Two groups of parameter estimates are included in the analysis: (a) parameter estimates across all SEWI simulation training cycles, indicating a robust model performance; (b) parameter estimates only after 500,000 training cycles in the SEWI simulation runs, indicating a model performance with emphasis on maximizing the information flows in modeling transitional effects in our landscape.

Fig. 9 plots the empirically obtained estimated parameters for location (x-axis) against scale (y-axis). Such a plot can help us select the best asymptotic form of the Bayes convergence factor using a dominance criterion, such as the *mean-variance-robustness* criterion. A desired probability distribution would have an estimated mean value closer to the 0.5 probability threshold (prevalence). Thus, estimated location parameters closer to 0.5 are dominant. On the other hand, we want our predicted probability distributions to minimize the level of uncertainty in our predictions. Thus, estimated scale parameters with smaller values are dominant. Finally, a desired probability distribution would have relative consistent estimated values of the location and scale parameters in both robust and informational assessments. We can see from Fig. 9 that the only asymmetric form of the Bayes convergence factor that meets all three dominance criteria is the one with $\alpha = 0.25$.

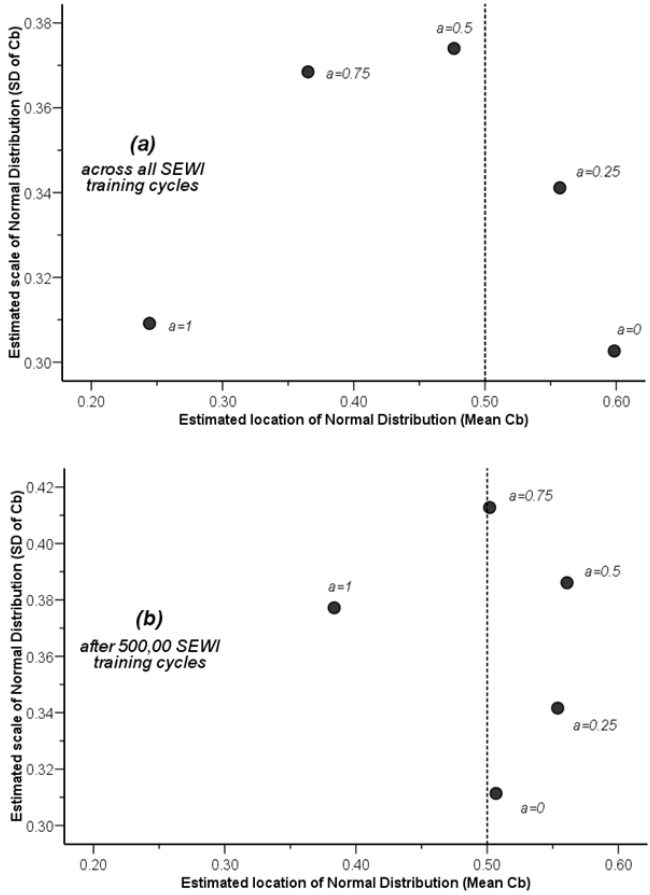

Fig. 9: Estimated location (μ) and scale (σ) parameters of selected asymmetric normal forms of the Bayes convergence factor from empirical data in the SEWI region: (a) robust estimates (across all training cycles); (b) maximum information estimates (after 500,000 cycles).

We can further enhance the quantitative assessment of the dominant asymmetric form of the $C_b$ metric, by computing explicitly the dominance criteria. The three dominance criteria can be combined as,

$f_N(\hat{\mu}_i, \hat{\sigma}_i) \succ f_N(\hat{\mu}_j, \hat{\sigma}_j)$ if and only if, $\forall m \neq k$:

$$\left(\frac{(0.5-\hat{\mu}_i)}{\hat{\sigma}_i}\right)_m - \left(\frac{(0.5-\hat{\mu}_i)}{\hat{\sigma}_i}\right)_k \geq \qquad (25)$$
$$\left(\frac{(0.5-\hat{\mu}_j)}{\hat{\sigma}_j}\right)_m - \left(\frac{(0.5-\hat{\mu}_j)}{\hat{\sigma}_j}\right)_k$$

where, the symbol "$\succ$" denotes dominant relationships, and,

$i, j$: unique combinations of location and scale (i.e., asymmetric forms of the $C_b$ metric).

$m, k$: unique groups for testing robustness (i.e., training cycle groupings).

$0.5 - \hat{\mu}_i \succ 0.5 - \hat{\mu}_j$: mean (location) criterion

$\hat{\sigma}_i \succ \hat{\sigma}_j$: variance (scale) criterion

$(\delta_m - \delta_k)_i \succ (\delta_m - \delta_k)_j$: robustness criterion, and $\delta$: any value or classification rule.

The results of the dominance criteria in the SEWI results visualized in Fig. 9 are summarized in Table 3. The values of the table cells represent the values of the differences in equation (25). The shaded cells signify the dominant asymmetric form of the Bayes convergence factor to be chosen.

Table 3: Table 3. Dominance values for assessing the selection of the Cb asymmetric normal form in the SEWI region.

|  |  | Robustness Groups | |
|---|---|---|---|
|  |  | All training cycles | 500,000 training cycles |
| Mean-Variance Groups | α = 0 | 1.071 | 0.071 |
|  | α = 0.25 | 16.744 | 15.744 |
|  | α = 0.5 | 0.286 | -0.714 |
|  | α = 0.75 | 0.987 | -0.013 |
|  | α = 1 | 1.596 | 0.596 |

Selecting the appropriate asymmetric form of the Bayes convergence factor allow us to infer additional information about the overall performance of our model. We can measure the deviation from a symmetric normal distribution (expected prior probabilities) that the estimated asymmetric form of the Bayesian convergence factor (observed posterior probabilities) yields. The P-P plots of this assessment are shown in Fig. 10. The thick curve represents the estimated cumulative probability distribution of the asymmetric $C_b$ predictive values observed in the SEWI region, and estimated from the simulation data. The estimated parameters (location, scale) are shown in the right side of each graph. The diagonal line represents the expected cumulative probability distribution of a symmetric distribution of predictive values (i.e., the expected predictive values at a prevalence threshold of 0.5). The parts of the predicted cumulative distribution curve that are above the expected one (diagonal) signify an increase in model accuracy that can be obtained from an asymmetric classification, while the parts of the predictive cumulative distribution curve below the expected diagonal line, signify a decrease in model accuracy. The point where the two lines meet (shown as the point of intersection of the reference lines), provide us with an estimated empirical prevalence level (threshold value for classification) that maximizes the modeling accuracy in our data. The net gain (or loss) in predictive value of our model due to the uncertainty in classification is the difference in the area that rests between the expected diagonal line, and the estimated observed curve.

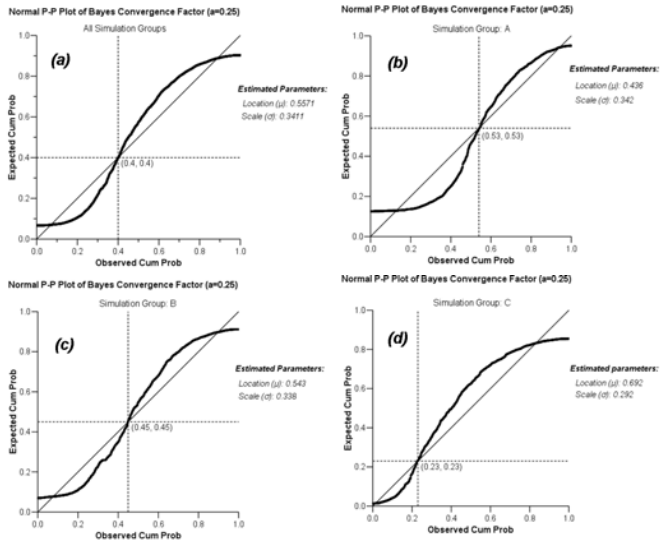

Fig. 10: Normal P-P plots for assessing Bayesian convergence simulation performance of the SEWI region: (a) across all simulation groups; (b) simulation group A; (c) simulation group B; (d) simulation group C.

From an initial observation of the models' accuracy within all simulation runs in the SEWI region, shown in sub-graph (a) of Fig. 10, the estimated prevalence threshold (=0.4) does not seem to deviate importantly from the expected one (0.5). Shifting the prevalence threshold would provide a 5.7% increase in the predictive value (informational gain) of the model. But, if we repeat our analysis for the SEWI regions' simulation groups (A, B and C), thus accounting for structural differences in the proportion of urban cells and exclusionary areas, we can see that spatial configuration affects considerably our actual model performance. For simulation group A, shown in sub-graph (b), the model performance is heavily dependent on the negative predictive values (estimated prevalence of 0.53 > 0.5), and produces poor overall model predictive values ($C_b$=0.436, or a 6.4% decrease in mean predictive value of the model). As the proportion of urban to exclusionary increases in the spatial composition of our simulation maps, the predictive value of the model increases substantially, and the estimated prevalence level decreases. Especially for group C, shown in sub-graph (d), a gain of 19.2% in model performance can be obtained from a shift in model prevalence (from 0.5 to 0.23).

## IV. DISCUSSION AND CONCLUSIONS

The analysis described above reveals the magnitude and multi-dimensionality of the spatial complexity involved in modeling land use change transitions in mixed and asymmetric landscapes in terms of amount and distribution of change. Performing spatial accuracy assessment requires the development and utilization of additional, advanced methods of assessment, related both to the models' predictive value in terms of quantity of change, but also to the performance of classifying the presence or absence of such a transition. It has been shown above that classification accuracy is closely related to the achieved modeling performance, and additional Bayesian metrics have been proposed, described and analyzed using the SEWI region case study. These advanced methods take into advantage the stochastic character of intelligent simulation models such as the LTM model, and can be used for performing model assessment in agent-based models of land use change, or other spatially explicit artificial intelligent modeling. The metrics described in this paper address different aspects of the spatial modeling performance such as assessing the predictive value of the model simulations (*PPV, NPV, DOR*), and estimating empirical convergence curves for enhancing classification accuracy ($C_b$). The proposed metrics and their assessment methodology allows the researcher and analyst to acquire a more holistic assessment of a models' spatial accuracy over space and time, especially in the presence of uncertainty about the transitional model thresholds.

The case study of the SEWI region used to illustrate the usage of the metrics, allow us to make assess the LTM model accuracy for simulating urban changes in the region. All metrics seem to confirm a general emergent model accuracy that appears to converge towards a 70% upper level. We can also see how the amount of urban change and exclusionary zones present in our landscapes dramatically affects the performance of the model. The latter result raises the significance of adjusting the classification prevalence threshold at spatially homogeneous scales in our simulation groups (e.g., implementing different thresholds for groups with different classes of urban change).

The results obtained also allow us to infer that in landscapes where the rate and amount of land use change vary importantly, symmetric spatial transition classification schemes are difficult to obtain. Instead we can enhance model predictions by assuming asymmetric spatial configurations, and by estimating the degree of asymmetry via a spatial *stochastic dominance* methodology. The practical significance of the proposed additional spatial model assessment metrics is that they can provide an "informational summary" of the simulated region or landscape ensembles. The use of the analysis and the performance of the metrics can help us in a multitude of ways. First, to understand and learn how well the model fits to different combinations of presence and absence of transitions in our landscapes, not simply how well the model fits our given data. Second, given that most spatial databases suffer from incomplete information and pre-simulation measurement errors, we can also derive (estimate) a theoretical accuracy that we would expect our model to assess, under the presence of such incomplete information data, and thus partially separate model from measurement errors in spatial simulations. Third, to understand the role and pattern of uncertainty in our simulations and model estimations. We can compare results across simulation runs (and thus quantitative patterns of change) that tend to provide less or more uncertain model performance, and understand the role of spatially-explicit patterns and cell configurations to model training and simulation. Fourth, to compare the significance or estimation contribution of transitional presence and absence (change versus no change) to our model performance, and the contribution of the spatial drivers and variables to the explanatory value of our model. Estimating model performance using different combinations of drivers (e.g., instead of groups A, B, C in the SEWI region, use of the same sampled boxes with different drivers, or using training sets with sequentially dropping a driver at a time), could allow us to estimate the differences in informational uncertainty for each driver combination or for single drivers within our simulations. Fifth, to compare measurements of informational uncertainty at different scales of spatial resolution. Pijanowski et al. (2003; 2005) showed the significance of using a scalable window for sensitivity analysis. Assessing model uncertainty of predictions for each of spatial resolutions can also enhance our knowledge about modeling at different spatial scales and selecting scales that produce lower uncertainty estimates.

Finally, the methodology and metrics developed in this paper allows for the development of a dynamic and adaptive modeling methodology. Beyond the aggregate level for which the assessment was performed for the purposes of this paper, it is both methodologically and computationally feasible to assess and adjust model accuracy at a simulation-to-simulation basis, in order to obtain dynamically enhanced simulation results. Especially in the case of agent-based modeling such a model assessment methodology can be inversed and iterated to obtain spatially robust and diverse future landscape configurations that optimize both the amount and degree of information contained in the simulation, and the emergence of stochastically dominant agent strategies.

ACKNOWLEDGMENT

Acknowledgments here.